# Automation of the Export Data from Open Journal Systems to the Russian Science Citation Index

Serhiy O. Semerikov[1][0000-0003-0789-0272], Vladyslav S. Pototskyi[1],
Kateryna I. Slovak[2][0000-0003-4012-8386], Svitlana M. Hryshchenko[2][0000-0003-4957-0904]
and Arnold E. Kiv[3]

[1] Kryvyi Rih State Pedagogical University, 54, Gagarina Ave., Kryvyi Rih, 50086, Ukraine
{semerikov, pototskiyvl}@gmail.com
[2] State Institution of Higher Education "Kryvyi Rih National University",
11, Vitali Matusevich St., Kryvyi Rih, 50027, Ukraine
slovak@fsgd.ccjournals.eu, s-grischenko@ukr.net
[3] Ben-Gurion University of the Negev, Beer Sheba, Israel
kiv@bgu.ac.il

**Abstract.** It is shown that the calculation of scientometric indicators of the scientist and also the scientific journal continues to be an actual problem nowadays. It is revealed that the leading scientometric databases have the capabilities of automated metadata collection from the scientific journal website by the use of specialized electronic document management systems, in particular Open Journal Systems. It is established that Open Journal Systems successfully exports metadata about an article from scientific journals to scientometric databases Scopus, Web of Science and Google Scholar. However, there is no standard method of export from Open Journal Systems to such scientometric databases as the Russian Science Citation Index and Index Copernicus, which determined the need for research. The *aim* of the study is to develop the plug-in to the Open Journal Systems for the export of data from this system to scientometric database Russian Science Citation Index. As a *result* of the study, an infological model for exporting metadata from Open Journal Systems to the Russian Science Citation Index was proposed. The SirenExpo plug-in was developed to export data from Open Journal Systems to the Russian Science Citation Index by the use of the Articulus release preparation system.

**Keywords:** Scientometric Indicators, Scientometric Databases, Specialized Systems for Electronic Workflow Support, SirenExpo plug-in.

## 1 Introduction

The formalized accounting of the scientist's productivity according to the published results is an important component of the evaluation of his activity, the activity of scientists and scientific institutions – is carried out with the help of scientometric databases. Internet-accessibility of the scientific publication for today is one of the top-priority requirements for its inclusion in any scientometric databases.

The main source of information about the publication is their annotations and other metadata posted on the website of the scientific journal. The use of standard protocols for metadata exchange promotes a better calculation of scientometric indicators not only of the scientist but also of the scientific journal itself (primarily its impact factor) [3; 5; 6; 8; 16].

Unfortunately, not all leading scientometric databases have the possibility of automated metadata collection from the scientific journal's site, which actualized the conduct of an appropriate study.

## 2    Literature Review and Problem Statement

The issue of qualitative and quantitative evaluation of published scientific results is due to the cause of the appearance of scientometry. Scientometrics determines the quality of scientific works and the quality of the scientist's work by analyzing scientific works on certain criteria.

One of the founders of scientometry is John Desmond Bernal, who described the laws of the functioning and development of science, the structure and dynamics of scientific activity, the interaction of science with the material and spiritual sphere of society, the role of scientometry in the social process in his work «The Social Function of Science» [1] of 1939.

After World War II, Derek John de Solla Price made a significant contribution to the development of science. Being a mathematician and a physicist, he defended his second thesis on the history of science. D. J. de Solla Price used quantitative methods to study science [13].

The term «scientometrics» was first used by V. V. Nalimov and Z. M. Mulchenko in the monograph «Scientometrics. Study of science as an information process», published in 1969. Authors define scientometrics as one of the branches of science, in which «science is viewed as a system that self-organizing and directs its own information flows» [10, p. 6]. «While studying science as an information process, it turns out to be possible to apply quantitative (statistical) research methods... It seems natural to call this direction of research – scientometrics» [10, p. 9].

A great contribution to scientometrics was made by Eugene Eli Garfield [6], who in 1960 founded Institute for Scientific Information. In 1964, E. Garfield launched the Science Citation Index [5], which became a powerful tool of scientometrics and became the basis of the scientometric database Web of Science.

The main scientometrics indicators are: Science Citation Index; $h$-index; $g$-index; $i$10-index and impact factor [14; 15].

Science Citation Index (SCI) – is a measure of the author's influence or scientific work on the development of science (see Fig. 1, Fig. 2). SCI reflects the total number of references to a particular scientific work or author in other scientific articles. The negative side of scientometric research using SCI is that this index does not take into account the time of article influence on science. That means the author, who created an article of poor quality about 20 years ago and quoted at least once a year, receives the same citation index as the good work that received 20 citations per 20 years.

Also, the citation index does not reflect the characterization of the scientific potential of the scientist. That is, a scientist who has written one work that has gained a certain popularity, and without having written more works, can have the same popularity with that scientist who has many scientific works. This and other shortcomings of the citation index prompted scientists to create new methods for assessing scientific papers.

**Fig. 1.** Indexed articles from SCI (according to [9, p. 3])

**Fig. 2.** The list of references to SCI (according to [9, p. 3])

The *h*-index was developed by Jorge E. Hirsch, a professor of physics at the California University of San Diego, who proposed «Hirsch's index» in 2005 [8], where he described the algorithm of the index, as well as the advantages and disadvantages of alternative methods (Table 1). According to J. Hirsch, the relationship between the *h*-index and the total number of citations can be described by the formula

$$N_{c,tot} = ah^2. \qquad (1)$$

J. Hirsch find empirically that *a* ranges between 3 and 5.

**Table 1.** Traditional methods for assessing the performance of a scientist according to [8]

| No | Method | Advantage | Disadvantage |
|---|---|---|---|
| (*i*) | total number of papers ($N_p$) | measures productivity | does not measure importance or impact of papers |
| (*ii*) | total number of citations ($N_{c,tot}$) | measures total impact | – hard to find and may be inflated by a small number of "big hits", which may not be representative of the individual if he or she is a coauthor with many others on those papers. In such cases, the relation in Eq. 1 will imply a very atypical value of *a* > 5;<br>– gives undue weight to highly cited review articles versus original research contributions. |
| (*iii*) | citations per paper ($N_{c,tot}/N_p$) | allows comparison of scientists of different ages | hard to find, rewards low productivity, and penalizes high productivity |
| (*iv*) | number of "significant papers", defined as the number of papers with more then *y* citations | eliminates the disadvantages of criteria (*i*), (*ii*) and (*iii*) and gives an idea of broad and sustained impact | *y* is arbitrary and will randomly favor or disfavor individuals, and y needs to be adjusted for different levels of seniority |
| (*v*) | number of citations to each of the *q* most-cited papers | overcomes many of the disadvantages of the criteria above | it is not a single number, making it more difficult to obtain and compare; also, *q* is arbitrary and will randomly favor and disfavor individuals |

*h*-index is a scientific metric that is a quantitative characteristic of the performance of a scientist, group of scientists or a country. According to J. Hirsch, the scientist has an index *h*, if his $N_p$ articles are quoted at least *h* times. Scientific works that do not satisfy this condition are not included in the indexation.

The peculiarity of the *h*-index is that it well reflects the results of scientific work when comparing the productivity of the scientific process in one area of activity. The disadvantage of the *h*-index is that the scientific index depends on the activity of the scientist. If a scientist ceases to engage in scientific work, his index will be the same as he was before, or at best, the scientist will have an *h*-index equal to the number of his articles.

The problem of staticity of the *h*-index was attempted to solve by a Belgian scientist from Universiteit Hasselt Leo Egghe by offering the *g*-index [4]. For a plurality of papers by a scholar sorted by the number of quotes, *g*-index is the largest number that *g* most cited articles received a total of at least $g^2$ citations (see Fig. 3).

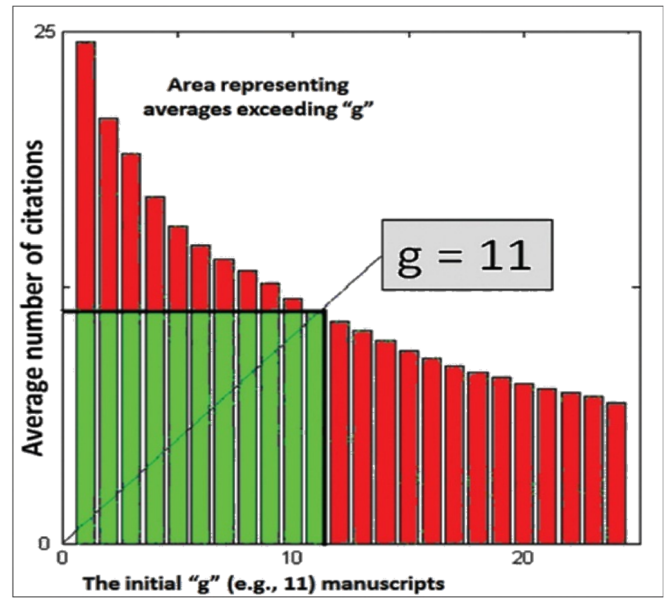

**Fig. 3.** The graph of the *g*-index (according to [14])

*i*10-index is the number of publications that were quoted not less than 10 times [7]. *i*10-index was developed by Google in 2011. This indicator depends predominantly on the age of the researcher and has a tendency to grow steadily. The five-year *i*10-index allows you to assess the current performance, and the overall – the impact of the work of a scientist on modern science without taking into account his past successes [8].

Impact factor (IF) is the ratio of the number of citations of articles of a certain journal to the total number of articles published in this journal (Eq. 2). In each particular year, the factor influencing the journal is the number of citations this year of the articles published in the journal over the past two years, divided by the total number of articles in this journal over the past two years [2].

$$IF_y = \frac{Citations_{y-1} + Citations_{y-2}}{Publications_{y-1} + Publications_{y-2}} \quad (2)$$

The basis for the analysis of the quantity and quality of the above indicators is the scientometric databases. They include bibliographic, abstract or full-text material on scientific publications, as well as tools for further tracking articles cited, internal search, etc. Scientometric databases are divided into commercial and free ones. The most popular commercial scientometric databases are Scopus and Web of Science.

Non-profit-oriented ones include Google Scholar, Russian Science Citation Index, DOAJ, WorldCat, Index Copernicus. The analysis of the leading scientometric databases has made it possible to identify their two main categories: 1) databases that index article's metadata automatically (Scopus, Web of Science), and 2) databases that article's metadata need to be entered by user's own hands (Russian Science Citation Index, Google Scholar and Index Copernicus).

Reduce the costs of supporting the work of the editorial board by creating the ability for members of the editorial board to work in the mode of remote access, increase the efficiency of editorial and publishing processes, improve scientific metrics, etc. provide specialized systems for supporting electronic document management.

The study identified four of the most popular systems that have different functionality for the publication of scientific papers – Open Journal Systems, DSpace, Koha and EPrints. The largest support for the editorial staff of the journal provides the Open Journal Systems (OJS) [12], the latest version of which (3.1) is partially documented and in a state of development. OJS is a free software developed by a nonprofit Public Knowledge Project. The system has a wide range of tools for editors of scientific journals. If some functionality is missing, it can be expanded using plug-ins. The OJS functionality and low system requirements have made it the standard to support the work of editorial boards of scientific journals. OJS successfully exports metadata about articles from scientific journals to such well-known scientometric databases as Scopus, Web of Science, Google Scholar, but there is no standard export method from OJS to such scientometric databases as Russian Science Citation Index and Index Copernicus. eLibrary development is used for submitting data to the Russian Science Citation Index that is the Articulus. The manual data input to Articulus is duplicated the work on preparing the description of the articles that has already was done in OJS, and therefore it's important to automate this process in order to reduce the unproductive time costs of members of the editorial board of the journal.

## 3   The Aim and Objectives of the Study

The aim of the study is to develop the plug-in to the Open Journal Systems for the export of data from this system to sciencemetric database Russian Science Citation Index.

To accomplish the set goal, the following tasks had to be solved:

1. to develop an infological model for the metadata export from Open Journal Systems to the Russian Science Citation Index;
2. to develop and test the plug-in to the Open Journal Systems for the metadata export to the Russian Science Citation Index.

# 4 Simulation and Development of Software for Export Automation from Open Journal Systems to Russian Science Citation Index

The OJS has a number of additions to export data in popular formats, as well as to the DOAJ open source directory. Unfortunately, with the transition to the new (third) version of OJS, the documentation for the plug-in developer is still not relevant.

In addition to the undocumented structure of the plug-in, there is another problem – the under-contentiousness of the metadata required for the Russian Science Citation Index. In Table 2, an infological model for the export of metadata from OJS to the Russian Science Citation Index was developed by analyzing the results of numerous experiments on the data export to/from the Russian Science Citation Index. As a result, XML structures were installed for import into the Russian Science Citation Index system.

**Table 2.** Model of Metadata Export from OJS to Russian Science Citation Index

| RSCI tag | Description |
| --- | --- |
| OperCard | a tag describing user information in the Articulus system (automatically filled in by the system when creating or importing a journal) |
| Titleid | log title identifier |
| ISSN | an international standard serial number that allows to identify periodicals |
| EISSN | an international standard serial number that allows to identify an electronic periodical |
| JournalInfo | the block where you can specify Title |
| Title (JournalInfo) | the subset of the JournalInfo tag, in which you can specify the name of the journal in different languages using the language attribute (lang="UKR", lang="ENG", etc). |
| Issue | the main tag, which describes all data of journal issue |
| Volume | journal volume |
| Number | issue number |
| AltNumber | end-to-end issue number |
| Part | part of issue |
| DateUni | date in YYYYMM format |
| IssTitle | volume name |
| Pages | number of pages in a volume |
| Articles | the main block, which contains a description of all the articles |
| Article | the block of the article, which describes all metadata articles |
| ArtType | type of article |
| Authors | the main block of the article's authors |

| RSCI tag | Description |
| --- | --- |
| Author | a block that describes single author using the tags: surname, initials, orgName (Organization Name), email, otherInfo (other information) |
| ArtTitles | article title block description. The block may include different languages that are specified when describing the title of an article |
| Text | text of the article |
| Codes | bibliographic description of the article, e. g, UDC, Dublin Core etc. |
| KeyWords | a block that describes the article keywords using the keyword tag |
| References | references to other articles |
| Files | files that belong to the article |

The development of the export model allowed us to move on to the next task – designing and developing a plug-in for export.

When developing the plug-in, the PHP programming language was used; the server with the LAMP-stack worked under Ubuntu Server 17.04. The following main assets were used to analyze and write the plug-in: PHPStorm, HeidiSQL, Git.

OJS has a number of shortcomings in the documentation that describes the rules and requirements for writing plug-ins to the system. But it is important to note that the program code is written by OJS authors, is well commented and uses comments in the form of Doxygen [17]. In the process of document search, the automatically generated Doxygen documentation for the OJS components was found.

To generate an XML file, you had to understand the structure of the import file to the Russian Science Citation Index – for this purpose, the Articulus system exported and identified the XML tags needed for export. Based on experiments with Articulus exports to the metadata described in Table 2, mandatory and optional fields were identified and their association with the OJS metadata was established.

The general scheme of the plug-in operation (see Fig. 4), is fairly transparent, which resulted in its rapid prototyping and development.

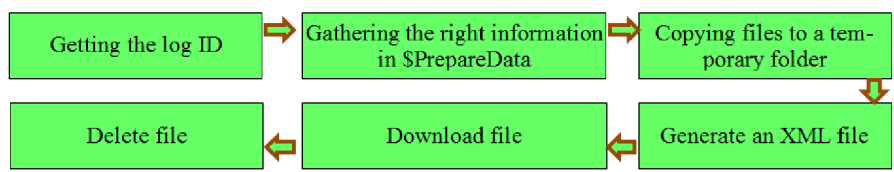

**Fig. 4.** General scheme of work of the developed plug-in

The created plug-in, called SirenExpo (https://github.com/Ladone/SirenExpo) (see Fig. 5). It was installed in the OJS system by copying the plug-in directory to the appropriate directory.

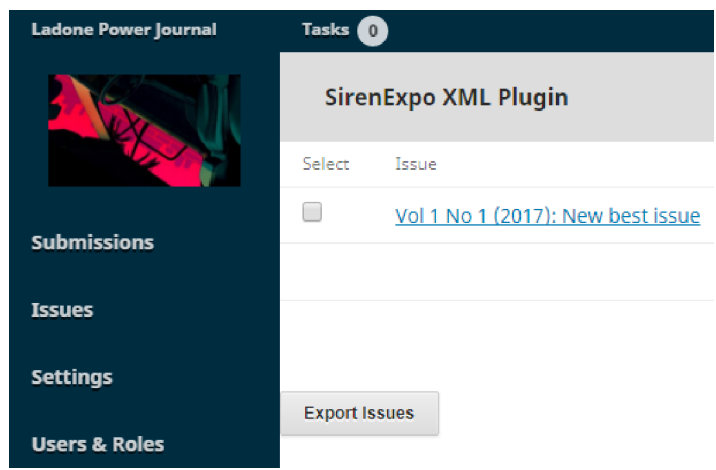

**Fig. 5.** SirenExpo plug-in interface

When using the plug-in, the user receives a list of journal issues that are available to him for export. To download the issue, select the issue and click on the "Export Issues" button. The program generates an archive and returns to the user. An example of the generated archive contains two pdf-files and one XML-file (see Fig. 6).

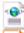

**Fig. 6.** Generated issue files

When authorizing the Articulus system, the user will receive a list of issues that can be transferred for indexing in the Russian Science Citation Index. In the menu, you need to click the "Restore Project" button. The user will go to the project recovery page, where you need to upload the archive generated by the SirenEXPO plug-in.

After uploading the file, the user receives a brief description of the journal issue, which was restored using the import function to Russian Science Citation Index (see Fig. 7).

At the Articulus, you need to click the "Restore project" button, and then a dialog box will open, in which you need to select the archive file generated by the SirenEXPO plug-in. After uploading the archive, all metadata required for the Russian Science Citation Index will be successfully imported. It is also possible to go to the restored project using the link "Open Project", in which there will be a window for editing the metadata of the journal. Editing of the restored project is depicted in Fig. 8.

So, with the help of SirenEXPO plug-in, you can export journal issues from the Open Journal Systems to the Russian Science Citation Index. The resulting archive is successfully uploaded into the Articulus – as a result, a new project with metadata imported from Open Journal Systems is created.

**Fig. 7.** Restoring of the project in Articulus

**Fig. 8.** Editing the restored project

## 5      Conclusions

As a result of the research, a new plug-in for the OJS was created, with the help of which you can export data to the scientometric database Russian Science Citation Index. The work describes the plug-in structure, the source code clearly shows where the information came from and how it was developed. If you need to create a new plug-in to OJS or add another scientometric database that needs to be imported, based on this development and research results, you can create new plug-ins for export to other scientometric databases, such as Index Copernicus.

## References

1. Bernal, J.D.: The Social Function of Science. George Routledge & sons Ltd., London (1939)
2. Cawkell, A.E.: Science perceived through the Science Citation Index. Endeavour. New Series. **1**(2), 57–62. doi:10.1016/0160-9327(77)90107-7
3. Directory of Open Access Journals. DOAJ. https://doaj.org/ (2018). Accessed 1 Feb 2018
4. Egghe, L.: Theory and practise of the g-index. Scientometrics. **69**(1), 131–152 (2006). doi:10.1007/s11192-006-0144-7
5. Garfield, E.: "Science Citation Index" – A New Dimension in Indexing Citation Indexes for Science: A New Dimension in Documentation through Association of Ideas. Science. **144**(3619), 649–654 (1964)
6. Garfield, E.: Citation Indexes for Science: A New Dimension in Documentation through Association of Ideas. Science. **122**(3159), 108–111 (1955)
7. Google Scholar Citations. Google Scholar Blog. https://scholar.googleblog.com/2011/07/google-scholar-citations.html (2011). Accessed 1 Feb 2018
8. Hirsh, J.E.: An index to quantify an individual's scientific research output. Proceedings of the National Academy of Sciences of the United States of America. **102**(46), 16569–16572 (2005). doi:10.1073/pnas.0507655102
9. Institute for Scientific Information: Science Citation Index 1964 Overview. Institute for Scientific Information, Philadelphia (1964)
10. Nalimov, V.V., Mul'chenko, Z.M.: Naukometriya. Izuchenie Razvitiya Nauki kak Informatsionnogo protsessa (Scientometrics. Study of science as an information process). Nauka, Moscow (1969) (in Russian)
11. Open Archives Initiative. Cornell University Library Information Technology, Ithaca. http://www.openarchives.org (2017). Accessed 1 Feb 2018
12. Open Journal Systems | Public Knowledge Project. Simon Fraser University Library, Burnaby. https://pkp.sfu.ca/ojs/ (2014). Accessed 1 Feb 2018
13. Price, D.: Little science, big science ... and beyond. Columbia University Press, New York (1986)
14. Ranjan, A., Kumar, R., Sinha, A., Nanda, S., Dave, K.A., Collette, M.D., Papadimos, T.J., Stawicki, S.P.: Competing for impact and prestige: Deciphering the "alphabet soup" of academic publications and faculty productivity metrics. International Journal of Academic Medicine. **2** (2), 187–202 (2016). doi:10.4103/2455-5568.196875
15. Samofal, O.I.: Naukometriia. Bibliometriia (Scientometrics. Bibliometry). Scientific Library of Yaroslav Mudryi National Law University, Kharkiv.

http://library.nlu.edu.ua/index.php?option=com_k2&view=itemlist&task=category&id=273 (2014). Accessed 1 Feb 2018
16. Tkachuk, V.V., Yechkalo, Yu.V., Semerikov, S.O.: Reitynh suchasnoho naukovtsia yak skladnyk reitynhu universytetu (The rating of a modern scientist as a component of the University's ranking). In: Proceedings of the International Scientific and Technical Conference on Development of Industry and Society, Kryvyi Rih National University, Kryvyi Rih, 24-26 May 2017
17. Van Heesch, D.: Doxygen: Generate documentation from source code. http://www.stack.nl/~dimitri/doxygen/ (2016). – Accessed 1 Feb 2018